\documentclass[aps,prl,twocolumn,superscriptaddress,showpacs,nobalancelastpage]{revtex4}
\usepackage{graphicx}
\usepackage{epstopdf}

\begin{document}


\title{Revealing the Supernova--Gamma-Ray Burst Connection with TeV
Neutrinos}

\author{Shin'ichiro Ando}
\affiliation{Department of Physics, School of Science, The University of
Tokyo, Tokyo 113-0033, Japan}
\affiliation{Department of Physics, The Ohio State University, Columbus,
OH 43210, USA}

\author{John F. Beacom}
\affiliation{Department of Physics, The Ohio State University, Columbus,
OH 43210, USA}
\affiliation{Department of Astronomy, The Ohio State University, Columbus,
OH 43210, USA}

\date{February 24, 2005; accepted June 21, 2005}

\begin{abstract}
Gamma-ray bursts (GRBs) are rare but powerful explosions displaying
 highly relativistic jets.
It has been suggested that a significant fraction of the much more
 frequent core-collapse supernovae are accompanied by comparably
 energetic but mildly relativistic jets, which would indicate an
 underlying supernova--GRB connection.
We calculate the neutrino spectra from the decays of pions and kaons
 produced in jets in supernovae, and show that the kaon contribution is
 dominant and provides a sharp break near 20 TeV, which is a sensitive
 probe of the conditions inside the jet.
For a supernova at 10 Mpc, 30 events above 100 GeV are expected in a 10
 s burst in the IceCube detector.
\end{abstract}


\maketitle


Long duration gamma-ray bursts (GRBs) have been found to be tightly
connected with core-collapse supernovae \cite{Hjorth03} (short duration
GRBs may be caused by another mechanism such as compact star mergers
\cite{Ando04b}).
Although many mysteries remain, this strongly indicates that the central
remnants of the core-collapse event, most likely black holes, drive the
observed relativistic (bulk Lorentz factor $\Gamma_b \agt 100$) jets of
GRBs.
High-energy neutrinos from accelerated protons in GRB internal shocks
are predicted to be detectable at future 1 km$^3$ \v Cerenkov detectors
such as IceCube \cite{Waxman97b}.
(If the protons escape and collide with external material, both the
delayed gamma rays from neutral pions and the neutrinos from charged pions
could be observable \cite{Katz94a}.)
The jet signature, but more mildly relativistic, may be common to
supernovae, which are much more frequent than GRBs, even
correcting for the effects of jet opening angle on observability.
If a significant fraction of core-collapse supernovae are accompanied by jets with
$\Gamma_b \sim 3$, perhaps $\sim 1\%$ according to late-time radio
observations \cite{Totani03}, then neutrinos could be the only prompt
signature of these hidden sources.

In addition to being more frequent, mildly relativistic jets are
expected to be much more baryon-rich (a ``dirty fireball'').
Both properties work quite positively for neutrino detectability.
Recently, Razzaque, M{\'e}sz{\'a}ros, and Waxman (RMW)
developed a model of high-energy (TeV) neutrino emission from jets
with $\Gamma_b \sim 3$ \cite{Razzaque04a}.
Because of the low Lorentz factor and high baryon
density, collisions among accelerated protons ($pp$) occur
efficiently, making pions that decay into neutrinos, and a nearby
core-collapse supernova at 3 Mpc is predicted to be detectable at
IceCube~\cite{Razzaque04a}.
If those high-energy neutrinos are detected, and correlated with an
optical supernova, then it would strongly and directly show the presence
of a mildly relativistic jet with significant kinetic energy, as well as
potentially being the first detection of extragalactic neutrinos.
Data from many supernovae would give the distribution of the jet Lorentz
factor, providing important insight into the supernova--GRB connection.

We extend the RMW model and significantly improve the detection
prospects; most importantly, we consider the kaon contribution, since compared to
pions, kaons have several advantages.
Due to their larger mass and shorter lifetime, kaons undergo less energy
loss before decaying into neutrinos.
This, and the relatively larger energy transferred to the daughter
neutrinos, means that the neutrino spectrum from kaons is harder and
more detectable than that from pions.
The neutrino spectrum from kaon decay has a sharp drop
near 20 TeV, which is a sensitive diagnostic of the acceleration
mechanism and the conditions inside the jet.
Since many detected events are expected (e.g., $\sim 30$ at 10 Mpc in
IceCube), our results greatly extend the range of detectability,
which increases the frequency as $\sim d^3$, where $d$ is the
distance of the furthest detectable objects.
While these models have many uncertainties, our results significantly
improve their detectability, such that even the existing AMANDA detector
can provide important constraints.


{\it Jet Dynamics.---}%
We briefly summarize the RMW model, using their notation
\cite{Razzaque04a}.
The jet kinetic energy is set to be $E_j = 3\times 10^{51}$ erg, which
is typical for GRBs.
Since we discuss a baryon-rich jet, we use the mildly relativistic value
for the  bulk Lorentz factor $\Gamma_b = 3$, and an assumed opening angle
of $\theta_j \sim \Gamma_b^{-1} = 0.3$.
By analogy with observed GRBs, it is natural to set the variability
timescale of the central object as $t_v = 0.1$ s.
The internal shocks due to shell collisions then occur at a radius $r_j
= 2\Gamma_b^2 c t_v = 5\times 10^{10}$ cm, smaller than a
typical stellar radius.
The comoving number densities of electrons and protons inside the jet
are $n_e^\prime = n_p^\prime = E_j/(2\pi \theta_j^2 r_j^2 \Gamma_b^2 m_p
c^3 t_j) = 4 \times 10^{20}$ cm$^{-3}$, where $t_j = 10$ s is the
typical GRB jet duration; the superscript $^\prime$ represents
quantities in the comoving frame of the jet.
It is assumed that fractions $\epsilon_e = \epsilon_B = 0.1$ of the jet
kinetic energy are converted into relativistic electrons and magnetic
fields, also by analogy to GRBs.
These electrons lose energy immediately by synchrotron radiation.
The synchrotron photons, however, thermalize because of the high optical
depth, $n_e^\prime \sigma_T \Gamma_b c t_v = 2\times 10^6$, where
$\sigma_T$ is the Thomson cross section, which makes the jets invisible
with prompt gamma rays, unlike GRBs.
As a result, the magnetic field strength, photon temperature and number
density are given by $B^\prime = [4\epsilon_B E_j / (\theta_j^2 r_j^2
\Gamma_b^2 c t_j)]^{1/2} = 10^9$ G, $E_\gamma^\prime = [15 (\hbar c)^3
\epsilon_e E_j / (2\pi^3 \theta_j^2 r_j^2 \Gamma_b^2 c t_j)]^{1/4} = 4$
keV, and $n_\gamma^\prime = E_j \epsilon_e / (2\pi \theta_j^2 r_j^2
\Gamma_b^2 c t_j E_\gamma^\prime) = 8 \times 10^{24}$ cm$^{-3}$.
It is assumed that the internal shocks accelerate protons with a spectrum
$\sim E_p^{-2}$, normalized to the jet total energy.
The maximum proton energy is obtained by comparing the proton
acceleration timescale $t_{\rm acc}^\prime \simeq E_p^\prime / eB^\prime
= 10^{-12} ~{\rm s} ~(E_p^\prime/1~{\rm GeV})$ \cite{Katz94b} with the
energy-loss timescales.
RMW assume that radiative cooling by the synchrotron and Bethe-Heitler
($p \gamma \to p e^+ e^-$) processes dominates, and that the maximum
proton energy is $2 \times 10^6$ GeV.


{\it Proton and Meson Cooling.---}%
We describe our extension of the RMW model, including the maximum
acceleration and meson cooling arguments.
At energies below the photopion production (hereafter $p\gamma$)
threshold $E_{p\gamma,{\rm th}}^\prime = 0.3/E_\gamma^\prime ~{\rm
GeV}^2 = 7\times 10^4$ GeV, the proton acceleration timescale is much
shorter than any energy-loss timescale.
Above the $p\gamma$ threshold, we find that the most competitive cooling
mechanism is the $p\gamma$ process itself, due to the very high photon
density.
Assuming 15\% energy lost from an incident proton in each $p\gamma$
interaction, $\Delta E_p^\prime = 0.15 E_p^\prime$, and using $\sigma
_{p\gamma}=  5 \times 10^{-28}$ cm$^2$ \cite{Eidelman04}, we obtain a
cooling timescale $t_{p\gamma}^\prime = E_p^\prime / (c \sigma_{p\gamma}
n_\gamma^\prime \Delta E_p^\prime) = 6\times 10^{-8}$ s.
Equating $t_{\rm acc}^\prime = t_{p\gamma}^\prime$, we  obtain
$E_p^\prime = 5\times 10^4$ GeV, slightly less than the
threshold energy of the $p\gamma$ interaction.
Thus when the $p\gamma$ interaction becomes
accessible, it prevents further acceleration, so we
take $E_{p,{\rm max}}^\prime = 7\times 10^4$ GeV, in contrast to RMW.
As discussed below, we focus on the break in the kaon-decay neutrino
spectrum as a direct observable of the maximum proton energy.

Accelerated protons produce mesons efficiently via $pp$ interactions,
since the optical depth is $\tau_{pp}^\prime = n_p^\prime \sigma_{pp}
\Gamma_b c t_v = 2\times 10^5$, where $\sigma_{pp} = 5\times 10^{-26}$
cm$^2$ \cite{Eidelman04}.
The meson multiplicity in each $pp$ interaction is taken to be 1 for
pions, and 0.1 for kaons; this is the right ratio \cite{Alner85}, but we
have been conservative in the normalization to focus on the most
energetic mesons.
We assume that the mesons are produced with 20\% of the parent proton
energy, so that they follow the original spectrum of accelerated protons
with power-law index $-2$.
They cool, however, by radiative (synchrotron radiation and inverse
Compton scattering off thermal photons) and hadronic ($\pi p$ and $Kp$)
processes.
The cooling timescales are $t_{rc}^\prime = 3m^4 c^3 / [4 \sigma_T m_e^2
E^\prime (U_\gamma^\prime + U_B^\prime)]$ and $t_{hc}^\prime = E^\prime
/ (c \sigma_h n_p^\prime \Delta E^\prime)$ for radiative and hadronic
cooling, where $m$ and $E^\prime$ are the meson ($\pi$ or $K$) mass and
energy, $U_\gamma^\prime = E_\gamma ^\prime n_\gamma^\prime$ and
$U_B^\prime = B^{\prime 2}/ 8\pi$ are the energy densities of photon and
magnetic fields, $\sigma_h$ is the cross section for meson-proton
collisions, and $\Delta E^\prime$ is the energy lost by the incident
meson in each collision.
Adopting $\Delta E^\prime = 0.8 E^\prime$ \cite{Brenner82} and $\sigma_h
= 5\times 10^{-26}$ cm$^2$ for pions and kaons \cite{Eidelman04},
$t_{hc}^\prime$ is energy independent, while $t_{rc} ^\prime \propto
E^{\prime -1}$.
Since the total cooling timescale is given by $t_c^{\prime -1} =
t_{hc}^{\prime -1} + t_{rc}^{\prime -1}$, it is dominated by hadronic
cooling at lower energies, and radiative cooling at higher energies.

If mesons decay faster than they cool, then the daughter neutrinos
maintain the spectrum shape; otherwise, the spectrum becomes steeper.
Below the cooling break energy $E_{cb}^{\prime (1)}$, obtained
by equating $\gamma^\prime \tau = t_c^\prime (\sim t_{hc} ^\prime)$,
where $\gamma^\prime$ and $\tau$ are the meson Lorentz factor and
proper lifetime, there is no suppression because mesons
immediately decay after production.
Above $E_{cb}^{\prime (1)}$, the cooling suppression factor is the
ratio of two timescales, i.e., $t_c^\prime/ \gamma^\prime\tau$.
When the energy-loss process is dominated by hadronic cooling,
the suppression factor is $\propto E^{\prime -1}$.
When it is dominated by radiative cooling,
above an energy $E_{cb}^{\prime (2)}$, it is
$\propto E^{\prime -2}$; we obtain $E_{cb}^{\prime (2)}$ by solving
$t_{hc}^\prime = t_{rc}^\prime$.
RMW neglected hadronic cooling of mesons; it is not a
large effect for pions, but it is important for kaons.

\begin{figure}
\begin{center}
\includegraphics[width=8.25cm]{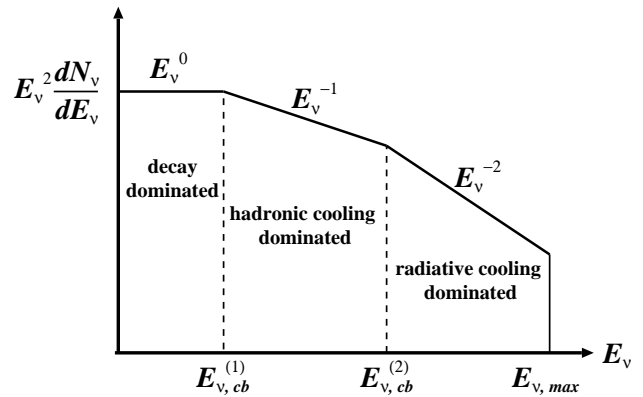}
\caption{Diagram for the neutrino spectrum from pion or kaon decays.
 The spectral break energies are given by Eqs. (\ref{eq:break energy 1
 for pion}) and (\ref{eq:break energy 2 for pion}) for pions, and by
 Eqs. (\ref{eq:break energy 1 for kaon})  and (\ref{eq:break energy 2
 for kaon}) for kaons. These spectral features are smeared out in
 detection  (see Fig. \ref{fig:event}).
\label{fig:cooling}}
\end{center}
\end{figure}


{\it Neutrino Spectrum.---}%
Charged pions and kaons, which are assumed to carry 20\% of the original
proton energy, decay into neutrinos through $\pi^\pm, K^\pm\to\mu^\pm +
\nu_\mu (\bar\nu_\mu)$ with branching ratios 100\% and 63\%.
The neutrino energy in the observer frame is related to the parent meson
energy in the jet rest frame as follows: $E_\nu = \Gamma_b E_\pi^\prime
/4$ and $E_\nu = \Gamma_b E_K^\prime /2$.
The Lorentz factor represents the Doppler boosting effect, and 4 and 2
in the denominators reflect the fraction of the parent energy conveyed
by the daughter neutrino in the case of pion and kaon decays.
Secondary neutrinos from muon decays are irrelevant since muons
immediately undergo radiative cooling~\cite{Razzaque04a}.
The energies and densities here are similar to the case of neutrino
production in Earth's atmosphere, which makes the calculation more
robust.

Figure \ref{fig:cooling} is a diagram for the neutrino spectrum from
meson decays, characterized by two spectral breaks and a maximum
neutrino energy.
For neutrinos from pion decay, we obtain the cooling break energies:
\begin{eqnarray}
E_{\nu,cb}^{\pi (1)}&=& 30~{\rm GeV},
 \label{eq:break energy 1 for pion}\\
E_{\nu,cb}^{\pi (2)}&=& 100 ~{\rm GeV},
 \label{eq:break energy 2 for pion}
\end{eqnarray}
corresponding to $E_{\pi,cb}^{\prime (1)}$ and $E_{\pi,cb}^{\prime
(2)}$.  The dependence on the jet parameters is given by $E_j^{-1}
\Gamma_b^7 \theta_j^2 t_j t_v^2$ and $(\epsilon_e + \epsilon_B)^{-1}
\Gamma_b$ for Eqs.~(\ref{eq:break energy 1 for pion}) and (\ref{eq:break
energy 2 for pion}), respectively.
We note that the first break energy is strongly sensitive to the value
of $\Gamma_b$ (it is less severe if one assumes
$\Gamma_b \sim \theta_j^{-1}$, following RMW).
This means that the model is quite uncertain, but at the same time, that the
detection of neutrinos could precisely constrain the Lorentz factor of the jet.

The neutrino spectrum from kaon decays is much more favorable, for
three reasons.
First, radiative cooling is much less efficient
than for pions, since kaons are heavier and the radiative cooling timescale
is $t_{rc}^\prime \propto m^4$.
Second, the kaon lifetime is a factor $\sim 2$ shorter.
Third, a larger mass also shortens the particle lifetime because of a
smaller Lorentz factor at fixed energy.
Thus the cooling breaks of kaons occur at much higher
energies:
\begin{eqnarray}
E_{\nu,cb}^{K (1)}&=& 200~{\rm GeV},
 \label{eq:break energy 1 for kaon}\\
E_{\nu,cb}^{K (2)}&=& 20,000 ~{\rm GeV},
 \label{eq:break energy 2 for kaon}
\end{eqnarray}
where the scaling is the same as Eqs. (\ref{eq:break energy 1 for pion})
and (\ref{eq:break energy 2 for pion}).
The maximum energy $E_{\nu,{\rm max}} = \Gamma_b E_{K,{\rm max}}^\prime
/ 2$ is only slightly above the second break for a canonical parameter
set, although this could be changed for other parameter choices.
Measurement of the sharp edge of the neutrino spectrum would be a
sensitive test of the maximum proton energy, and hence the physical
conditions in the jet.


{\it Neutrino Burst Detection.---}%
We first estimate the normalization of the neutrino spectrum,
evaluating the fluence at the first break energy, $F_{\nu,0}
\equiv F_\nu (E_{\nu,cb}^{(1)})$.
Assuming efficient energy conversion from protons to mesons, and that
half of the mesons are charged, we obtain
\begin{equation}
F_{\nu,0} = \frac{\langle n\rangle B_\nu}{8}\frac{E_j}{2\pi \theta_j^2 d^2
  \ln (E_{p,{\rm max}}^\prime/E_{p,{\rm min}}^\prime)}
  \frac{1}{E_{\nu,cb}^{(1)2}},
\label{eq:fluence}
\end{equation}
where $d$ is the source distance, $\langle n \rangle$ is the
meson multiplicity (1 for pions and 0.1 for kaons), $B_\nu$ is the
branching ratio of the decay into neutrino mode (1 for pions and 0.6 for
kaons), and the factor $\ln (E_{p,{\rm max}}^\prime/ E_{p,{\rm
min}}^\prime)$ normalizes the proton spectrum to the jet energy.
For canonical parameter choices and for a nearby source at  $d = 10$ Mpc,
$F_{\nu,0}$ becomes $5\times 10^{-2}$ and $5\times 10^{-5}$ GeV$^{-1}$
cm$^{-2}$, for neutrinos from pion and kaon decays, respectively.
The parameter dependence is $E_j^3 \Gamma_b^{-14} \theta_j^{-6}
t_j^{-2} t_v^{-4} d^{-2}$.

\begin{figure}
\begin{center}
\includegraphics[width=8.25cm]{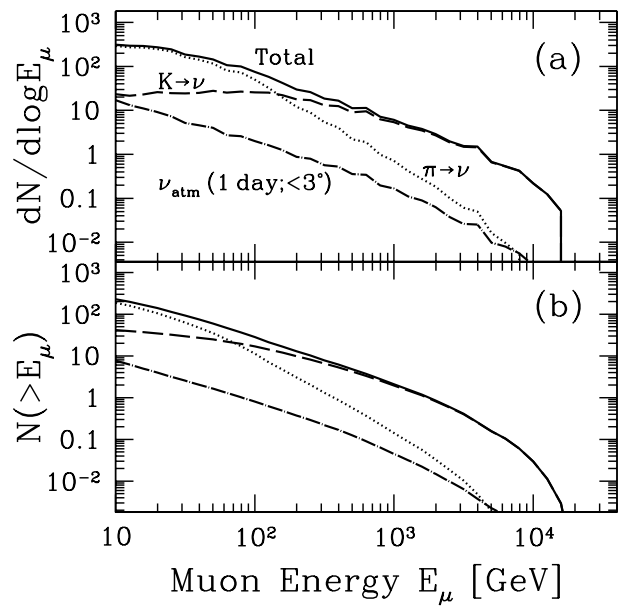}
\caption{(a) Event spectrum of neutrino-produced muons from a supernova
 at 10 Mpc in a 1 km$^3$ detector. Contributions from $\pi^\pm$ and
 $K^\pm$ decays are shown as dotted and dashed curves, and the total as
 a solid curve.  The atmospheric neutrino background is shown for
 comparison; it is evaluated for 1 day and within a circle of 3$^\circ$
 radius. (b) The same, but cumulative event number above a given
 energy. \label{fig:event}}
\end{center}
\vspace{-0.5cm}
\end{figure}

We calculated the expected signal from one supernova neutrino burst,
using the code ANIS (All Neutrino Interaction Generator)
\cite{Gazizov04}.
We neglect the effects of neutrino oscillations, as they are below the
uncertainties of the model.
Figure \ref{fig:event}(a) shows the event spectrum from the muon neutrinos
and antineutrinos from a supernova at 10 Mpc, and in
Fig. \ref{fig:event}(b), we show the yields above a given energy.
We used a detector effective area of 1 km$^2$, which is reasonable for
IceCube in the case of upgoing muons \cite{Ahrens04}.
We took into account the muon range, which effectively
enlarges the detector volume, and evaluated the muon energy when it
enters the detector if it is produced outside, or at the production
point otherwise.
Since the spectrum of neutrinos from pions falls steeply, their
expected event spectrum is also steep, and therefore, if we lower the
threshold, many more events would be expected, as shown in
Fig. \ref{fig:event}(b).
The spectrum of neutrinos from kaons, on the other hand, is much
flatter, making them the dominant component at high energies.

If we take 100 GeV as the threshold muon energy for IceCube (for a
transient point source), we expect about 30 events from a core-collapse
supernova at 10 Mpc, mostly from kaons.  (If the proton spectral
index is not $-2.0$, but is instead $-1.5$ or $-2.5$, the expected number of events
is 40 or 3, respectively.)
These events cluster in a 10 s time bin and a $\sim 3^\circ$ angular bin
surrounding the supernova, which allows very strong rejection of
atmospheric neutrino backgrounds.
If the source is farther and the expected number only a few events, then
we may use a more conservative time bin, e.g., a 1-day bin correlated
with optical observations, considering the time uncertainty between the
neutrino burst and an optical supernova.
The atmospheric neutrino background for 1 day and
within a circle of $3^\circ$ radius is also shown in
Fig. \ref{fig:event}.
The spectral break corresponding to the maximum proton energy
might be detectable around $E_\mu \sim 20$ TeV,
if the source is very close or the jet energy $E_j$
larger than assumed.
Recalling that the proton acceleration is limited by the $p\gamma$
interaction, this probes the photon number density in the jet, an essential
quantity of the model.
If the neutrinos come from below the detector,
it may be possible to reduce the detector threshold, so that the
large yield due to pion decays could be detectable too.


{\it Discussion and Conclusions.---}%
The RMW model, while speculative, is an specific and intriguing proposal
for a common thread connecting GRBs and core-collapse supernovae, namely
the presence of jets with energies around $3 \times 10^{51}$ erg.
For GRBs, which are rare events, the jets would be highly relativistic
and revealed; for supernovae, which are frequent, the jets would be
mildly relativistic and hidden, except from neutrino telescopes.
The most basic features of the RMW model are a proton spectrum falling
as $E_p^{-2}$, carrying most of the jet energy, and a suitable target
density for neutrino production.
In analogy to the observed properties of GRBs, and evidence for a
supernova--GRB connection, these requirements do not seem unreasonable.
What is significant is that these considerations can be directly and
easily constrained with neutrino detectors, lessening the dependence on
theory.

The key remaining issue is the fraction of supernovae which have these jets.
Some type Ic supernovae do suggest the presence of mildly relativistic jets,
based on their late-time radio emission; the fraction of all core-collapse supernovae
with jets is perhaps $\sim 1\%$~\cite{Totani03}, significantly larger than the
fraction of supernovae with highly relativistic jets, i.e., GRBs.
It is possible that mildly relativistic jets are in fact much more common, but
are completely choked in the hydrogen envelopes of type II supernovae
(RMW assumed that nearly all supernovae have jets).
If such mildly relativistic jets accompanied SN 1987A and were directed
toward Earth, the Kamiokande-II and IMB detectors would have seen many events,
the small detector sizes being more than compensated by the closer distance.
This suggests that such jets did not exist in SN 1987A; additionally, it was dark
in the radio~\cite{Ball01}.

In any case, the nearby core-collapse supernova rate is high enough to allow
testing these models soon, especially if type II supernovae have jets which can
only be revealed by neutrinos.
Within 10 Mpc, the rate of core-collapse supernovae is more
than 1 yr$^{-1}$, with a large contribution from galaxies around
3--4 Mpc~\cite{Ando05}.
At 3 Mpc, we expect about 300 events in a 10 s burst in IceCube, which
would be very dramatic.
The electron neutrino
shower channel, where the detected energy more faithfully reveals the
neutrino energy, may be especially helpful~\cite{Beacom04c}.
Even in the already-operational AMANDA, though the effective area is
smaller than that of IceCube, we expect $\sim 10$ events.
Since these events would arrive within 10 s, from a specific point
source, the background rejection should be excellent, and AMANDA could
already constrain these models.

At larger distances, the galaxy distribution is smooth enough to allow simple
scaling: the number of sources increases as $d^3$, while the
neutrino signal per supernova decreases as $d^{-2}$.  For example, at 20 Mpc,
the expected number of neutrino events in IceCube is still several,
and the total supernova rate is greater than about 10 yr$^{-1}$.
That distance contains the Virgo cluster (which increases the supernova rate
somewhat beyond the simple scaling) in the Northern hemisphere, which is an
attractive target for South Pole neutrino telescopes.   The effect of
beaming is to reduce the source frequency at a given distance, but to increase
the source fluence in a compensating way.  Given the large assumed opening
angle, the probability of having the jets directed towards Earth is relatively large,
around $\sim 10\%$.
Even taking into account that perhaps not all supernovae have jets, this
model predicts rich detection prospects for IceCube.
The detection of a prompt burst of high-energy neutrinos would
reveal the time and direction of the core collapse event, which would be
useful for forecasting the optical supernova, and for searching for a
gravitational wave signal in coincidence.


\medskip
\begin{acknowledgments}
We are grateful to Peter M{\'e}sz{\'a}ros, Soebur Razzaque, and Eli
 Waxman for extensive discussions, and to Steve Barwick,
 Juli{\'a}n Candia, and Hasan Y{\"u}ksel for comments.
This work was supported by The Ohio State University, and
S.A. by a Grant-in-Aid for JSPS Fellows.
\end{acknowledgments}



\end{document}